\documentclass{ifacconf}

\usepackage{graphicx}      
\usepackage{natbib}        
\usepackage{amsmath}
\usepackage{graphicx}
\usepackage{svg}
\usepackage{amsmath,amssymb,amsfonts}
\usepackage{textcomp}
\usepackage{xcolor}
\usepackage{xcolor,soul}
\usepackage{array}
\usepackage{times}
\usepackage{comment}
\begin{document}
\begin{frontmatter}

\title{Consensus of hierarchical multi-agent systems with a time-varying set of active agents} 

\thanks[footnoteinfo]{This research was funded by the NEON (New Energy and mobility Outlook for the Netherlands) Crossover NWO (Dutch Research Council) Grant, project number 17628.}

\author[First]{Victor Daniel Reyes Dreke} 
\author[First]{Mircea Lazar} 

\address[First]{ Control Systems Group, Department of Electrical Engineering, Eindhoven University of Technology, The Netherland (e-mail: v.d.reyes.dreke@tue.nl, m.lazar@tue.nl).}

\begin{abstract}                
Time-varying hierarchical multi-agent systems are common in many applications. A well-known solution to control these systems is to use state feedback controllers that depend on the adjacency matrix to reach consensus. This solution has been applied so far to multi-agent systems with fixed or time-varying communication topologies. In this paper, we consider single-integrator multi-agent systems where only a subset of the agents are active at any given time and the set of active agents is time-varying. This type of multi-agent system is relevant in applications such as modular multilevel converters and water pumping systems. We develop a switching algorithm that periodically selects a set of active agents through a set of graphs that connect the follower agents with a leader agent.  We further prove that the developed switching algorithm combined with a classic consensus protocol yields convergence to a consensus state defined by the leader despite a time-varying set of active agents. The effectiveness of the switching algorithm is illustrated in two benchmark examples: a modular multilevel converter and a water pumping system.
\end{abstract}


\end{frontmatter}

\section{INTRODUCTION}
\label{sec:Introduction}
Hierarchical multi-agent systems (H-MAS) are playing an important role in many applications, for example in modular multilevel converters (MMCs), microgrids, and water pumping systems. 
In the above applications, a leader such as the grid operator coordinates the control actions of the remaining agents of the system (followers) to achieve a common goal (a consensus value) such as a certain power demand. However, calculating and coordinating the control actions is a challenge that increases in complexity depending on the common goal and system characteristics. Further, the complexity of this task also increases if the H-MAS has a time-varying topology or it is subject to constraints such as having a limited number of active agents.


Commonly, the classic consensus problem is solved using a static state feedback controller where the agent dynamics are single integrators and the network topologies are time-invariant \citep{Lunze2019}. In this case, a necessary condition for reaching consensus is the existence of a spanning tree in the communication network. The importance of the spanning tree is due to the fact that the state feedback controller is built based on the adjacency matrix that describes the communication between agents; see, for example, \citep{Morita2018, JIANG2022, CHENG2015112}. Moreover, in \citep{Stoorvogel2011}, the findings of \citep{Lunze2019} were extended to identical linear time-invariant (LTI) multi-input and multi-output (MIMO) systems.

The previous results were constrained to time-invariant communication, yet time-varying communication networks are equally important. Addressing this topic, \cite{Jadbabaie2003}  and \cite{Ren2005} offered a proof for reaching consensus for multi-agent systems under dynamically changing interaction topologies. {In \citep{Ren2005}, it is proven that a necessary condition for reaching consensus in time-varying communication networks with a fixed number of agents is that the union of the directed graphs representing the interconnection among agents has a spanning tree.} 
The finding from \citep{Ren2005} was successfully extended \citep{Mondal2022} to agents with nonlinear dynamics by using nonlinear dynamic inversion. In \citep{Li2010}, the authors investigated the consensus problem in uncertain communication environments and designed a distributed stochastic approximation consensus protocol. Similarly, as in \citep{Ren2005}, the necessary condition for achieving consensus is that the union of graphs has a spanning tree. 

In all the above-mentioned works, the results are mainly focused on reaching an average consensus in a leaderless communication network where the number of agents is fixed. 
However, in the case of H-MAS, the communication networks typically consider a leader-follower architecture. In this case, the dynamics of the leader are crucial for reaching consensus. From \citep{Shao2018}, it is known that if the leader contains self-loops, stability problems can arise. Additionally, these authors investigated the impact of hierarchical topologies and followers’ self-loops on the convergence performance of leader–follower consensus, including the convergence rate and robustness to switching topologies. A similar  study  was done in \citep{Shokri2020}, where stochastic agents were considered. Different from the above-mentioned examples, \cite{Shokri2020} use a gossiping communication algorithm that limits the communication per sampling time to only one agent. 

The above surveyed works consider H-MAS with a switching, time-varying topology,  but with a fixed set of active agents. However, as it is the case with, e.g., MMCs and water pumping systems, the set of active agents can change over time. In these applications, controlling the amount of active agents can improve the lifespans of agents or help fulfill a common goal. Therefore it is of interest to investigate if consensus can be reached in a H-MAS with a leader-follower architecture and with a time-varying set of active agents.

In this paper, we will refer to the above types of systems as time-varying hierarchical multi-agent systems (TH-MAS). For reaching consensus in a TH-MAS, we propose a switching algorithm that will coordinate the control actions by determining which followers should be active at any discrete time instant. Moreover, based on fundamental results from \citep{Lunze2019} and \citep{Ren2005}, we prove that under the newly developed switching algorithm, the states of all follower agents reach consensus. For evaluating the results of the proposed method, we tested the proposed control algorithm in two simulation examples describing a modular multilevel converter as in \citep{Ygor2021} and a water pumping system as in \citep{Kusumawardana2019}.

The remainder of the paper is organized as follows. In Section~\ref{sec:Preliminaries and System Description} useful notations and description of the studied TH-MAS system are presented. Section~\ref{sec:consensus_general} and \ref{sec:Case_of_Study} present the main results of the paper and the application-inspired examples, respectively. Section 5 summarizes the conclusions of the paper. 

\section{Preliminaries and System Description}
\label{sec:Preliminaries and System Description}
\subsection{Basic preliminaries}
\label{subsec:PRELIMINARIES}
Let $\mathbb{N}$, $\mathbb{N}_+$, $\mathbb{R}$, $\mathbb{R}_{+}$, and $\mathbb{Z}$ denote the set of natural numbers, the set of natural numbers without including 0, the field of real numbers, the field of non-negative real numbers and the set of integer numbers, respectively. For a set $\mathbb{S}\subseteq\mathbb{R}^n$ define $\mathbb{S}_{[a,b]}:=\{s\in\mathbb{S} \ : \ a\leq s\leq b\}$. Define an $n$-dimensional vector filled with ones as $\mathbf{1}_{n}$, and one filled with zeros as $\mathbf{0}_{n}$. The identity matrix is denoted as $\mathbf{I}_{n}$. The relations $>, \leq, <, \geq$ are applied element-wise for vectors and matrices. For example, let 
\begin{equation}
A=\left[\begin{array}{ll}
a_{11} & a_{12} \\
a_{21} & a_{22}
\end{array}\right] \text { and } B=\left[\begin{array}{ll}
b_{11} & b_{12} \\
b_{21} & b_{22}
\end{array}\right]
\end{equation}
then $A>(\geq)B$, if ${{a}_{11}} >(\geq){{b}_{11}}$, ${{a}_{12}}>(\geq){{b}_{12}}$, ${{a}_{21}}>(\geq){{b}_{21}}$ and ${{a}_{22}}>(\geq){{b}_{22}}$.

For a graph $\mathcal{G}$, the matrix representations, such as the adjacency, degree and Laplacian matrices are obtained using the methodology from \citep{Lunze2019}.
The adjacency matrix $\mathcal{A}$ is generated by a function $f_{\mathcal{A}}: \mathcal{G} \rightarrow \mathcal{A}$.
The degree matrix $\mathcal{D}$ is built as a diagonal matrix using a function $f_{\mathcal{D}}: \mathcal{G} \rightarrow \mathcal{D}$.
Finally, the corresponding Laplacian matrix $\mathcal{L}$ is calculated as 
\begin{equation}
    \label{eq:Laplacian_definition}
    \mathcal{L} =  \mathcal{D} - \mathcal{A}, \quad \mathcal{L} \in \mathbb{Z}^{(N+1)\times (N+1)}.
\end{equation}



\subsection{TH-MAS: System description}
\label{sec:MODEL_DESCRIPTION}

\begin{figure}[htbp]
\centerline{\includegraphics[width=7cm]{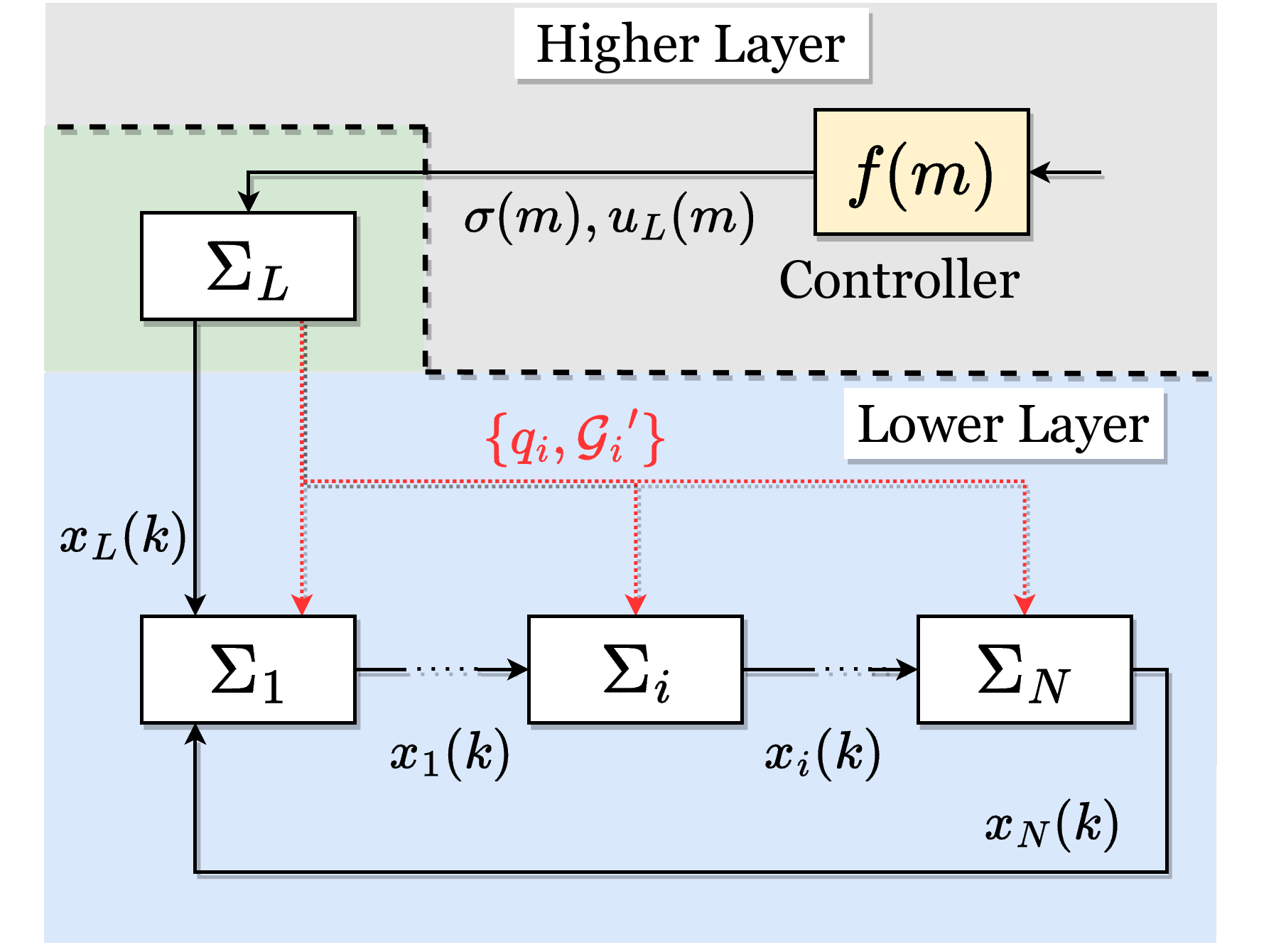}}
\caption{TH-MAS illustration.}
\label{Fig:H_MAS}
\end{figure}

Consider a time-varying hierarchical mutli-agent system (see Figure~\ref{Fig:H_MAS})  where, in the higher layer,  a discrete-time controller calculates the number of active followers, as $\sigma(m) \in \mathbb{N}_+$, and the states of the leader $({\Sigma}_{L})$ determined by its control input $(u_L(m) \in \mathbb{R})$, i.e.
\begin{equation}
    \label{eq:higher_layer_controller}
    f(m): \mathbb{R}^{n_f}  \rightarrow (\mathbb{N}_+,\mathbb{R}),
\end{equation}
where $n_f \in \mathbb{N}_+$ is the controller input dimension. 

The lower layer consists of $N$ follower agents which are labeled by $1,2,\ldots, N$ and a leader labeled by $N+1$.
The follower agent dynamics are defined as 
\begin{equation}
    \label{eq:follower_dynamics}
    \Sigma_{i} : \; {x}_{i} (k+1) = {x}_{i}(k) + w {q}_{i}(k) {u}_{i}(k), 
\end{equation}
where $w \in \mathbb{R}$, ${x}_{i}\;  \text{and}\; {u}_{i} \in \mathbb{R}$ are the  $i^{th}$ follower state and control input, respectively and $q_i \in \{0,1\}$ is a selection variable determining which agents are controlled at each time instant $k$. 

The leader states depend on the control input ($u_L(m)$) from the higher layer controller, such that  
\begin{equation}
    \label{eq:leader_dynamics}
    {x}_{L}(k+1) = (1-q_L(k))x_L(k) + {q}_{L}(k){u}_{L}(k),   
\end{equation}
where $m \in \mathbb{N}$ and $k \in \mathbb{N}$ are the discrete-time indices of the higher and lower layer, and
\begin{equation}
\label{eq:selection-of_the_leader}
q_{L}(k)= \begin{cases}1 & \text { if } k =mM,\\ 0 & \text { if }  k\in \left[ (m-1)M +1, \; mM \right),\end{cases} \;  
\end{equation}
with $M \in \mathbb{N}_+$ and for notation simplicity, $L = N+1$ labels the leader. The leader dynamics is interpreted as a piecewise constant function determined by the higher control level, since $x_L(0) =u_L(0)$.

The time-varying set of active  followers is expressed as an equality constraint, i.e. 
\begin{equation}
    \label{eq:equality_constraints}
    \sum^{N}_{i=1} {{q}_{i}(k)} = \sigma(m), \quad \forall k= mM.
\end{equation}



Hence, until the next update, it holds that: 
\begin{equation}
    \label{eq:piecewise_constraints}
    \sum^{N}_{i=1} {q}_{i}(k+1) = \sum^{N}_{i=1} {q}_{i}(k)    
\end{equation}
for all  $k\in \{ (m-1)M +1, \; mM\}$, while keeping 
$$ 1 \leq \sum^{N}_{i=1} {q}_{i}(k) \leq N, \quad \forall k \in \mathbb{N}.$$
\begin{figure}[tb]
\centerline{\includegraphics[width=7cm]{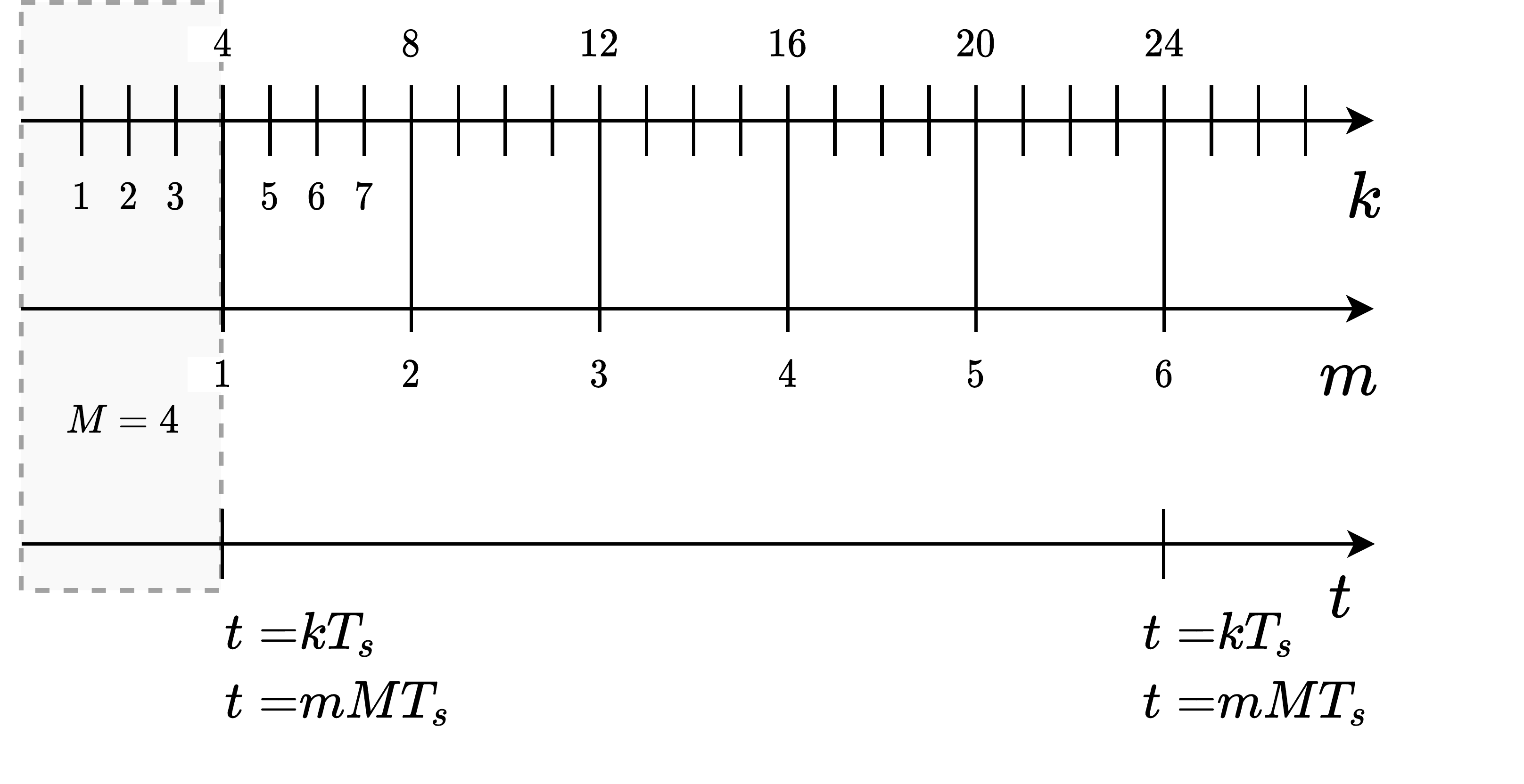}}
\caption{Illustration of the time relation between higher and lower hierarchical control layers.}
\label{Fig:Time_Rule}
\end{figure}
The described TH-MAS system can be characterized  as a multi-rate system with a higher and lower layer working at different sampling times as in Figure~\ref{Fig:Time_Rule}, such that the time ($t \in \mathbb{N_+}$) is quantized as
$$t =  k T_s = m {T}^{'}_{s},  $$
where $T_s \in \mathbb{R}$ and ${T}^{'}_{s} \in \mathbb{R}$ are their respective sampling times.
The sampling time difference is determined by
\begin{equation}
    \label{eq:time_difference}
    {T}^{'}_{s} =  M \cdot T_s  \;\; \text{with} \; M\in \mathbb{N}_+.
\end{equation}

Finally, an overall system is defined as 
\begin{equation}
    \label{eq:overall_model_discrete_2}
    {\Sigma}_d:= {\boldsymbol{x}}(k+1) = { \left[  \begin{array}{cc}
         \mathbf{I}_{N} &  \mathbf{0}_{N}\\
         \mathbf{0}^{\top}_{N}&  (1-q_L(k))
    \end{array}  \right]}\boldsymbol{x}(k) + B(k) \boldsymbol{u}(k), 
\end{equation}
with 
$$ B(k) = { \left[\begin{array}{cccc}
w q_1(k) & 0 & \cdots & 0 \\
0 & w q_2(k) & 0 & \vdots \\
\vdots & 0 & \ddots & \vdots \\
0 & \cdots & \cdots & q_L(k)
\end{array}\right]} \; \text{and} \; \boldsymbol{x_0} = \boldsymbol{x}(0), $$
where the state and control input vectors are $$\boldsymbol{x}(k) := \left[x_1(k),\ldots,x_N(k),x_L(k)\right]^{\top}\in 
\mathbb{R}^{N+1}$$ and $$\boldsymbol{u}(k) := \left[u_1(k),\ldots,u_N(k),u_L(k)\right]^{\top}\in 
\mathbb{R}^{N+1}.$$ 

\subsection{Communication Network for TH-MAS}
A directed graph ${\mathcal{G}} = ({\mathcal{V}} ,{\mathcal{E}})$ represents the communication network of the TH-MAS.  The corresponding set of nodes is ${\mathcal{V}}=\{1,2, \ldots, N+1\}$ with a set of directed edges ${\mathcal{E}} \subseteq {\mathcal{V}} \times {\mathcal{V}}$. 
Since the communication network has only one leader, the communication network that only contains the follower agents is represented by the graph $${\mathcal{G}^{'}} = ({\mathcal{V}^{'}} ,{\mathcal{E}^{'}}) \; \text{with} \; {\mathcal{V}^{'}} \subseteq {\mathcal{V}} \; \text{and} \; {\mathcal{E}^{'}} \subseteq {\mathcal{E}}.$$ 

Notice that the system constraint \eqref{eq:equality_constraints}  results in a time-varying graph for the followers network ${\mathcal{G}^{'}(k)}$ where the set of node cardinality changes, such that 
\begin{equation}
    \label{eq:set_of_node_followers_tv}
    |{\mathcal{V}^{'}}(k)| = {\sigma(m)}, \; \forall k = m M 
\end{equation}
and, therefore,
\begin{equation}
    \label{eq:set_of_node_followers_cardinality}
    |{\mathcal{V}^{'}}(k+1)| = |{\mathcal{V}^{'}}(k)|, 
\end{equation}
for all  $k\in \{ (m-1)M +1, \; mM\}$. 

\begin{rem}
\label{rmk:spanning_tree_}
A directed graph has a spanning tree if there exists a node (agent), which is called the root
node such that there exist directed paths from the root to any node (agent) in the graph. This is a necessary condition for all the agents reaching consensus. Hence, in this paper, both ${\mathcal{G}}$ and ${\mathcal{G}^{'}(k)}$ have a spanning tree with the leader $\Sigma_{L}$ as the root. 
\end{rem}

\section{Consensus of TH-MAS}
\label{sec:consensus_general}

\subsection{Consensus of TH-MAS: Higher Layer}
\label{subsec:Switching_algorithm}
\begin{figure}[htbp]
\centerline{\includegraphics[width=8cm]{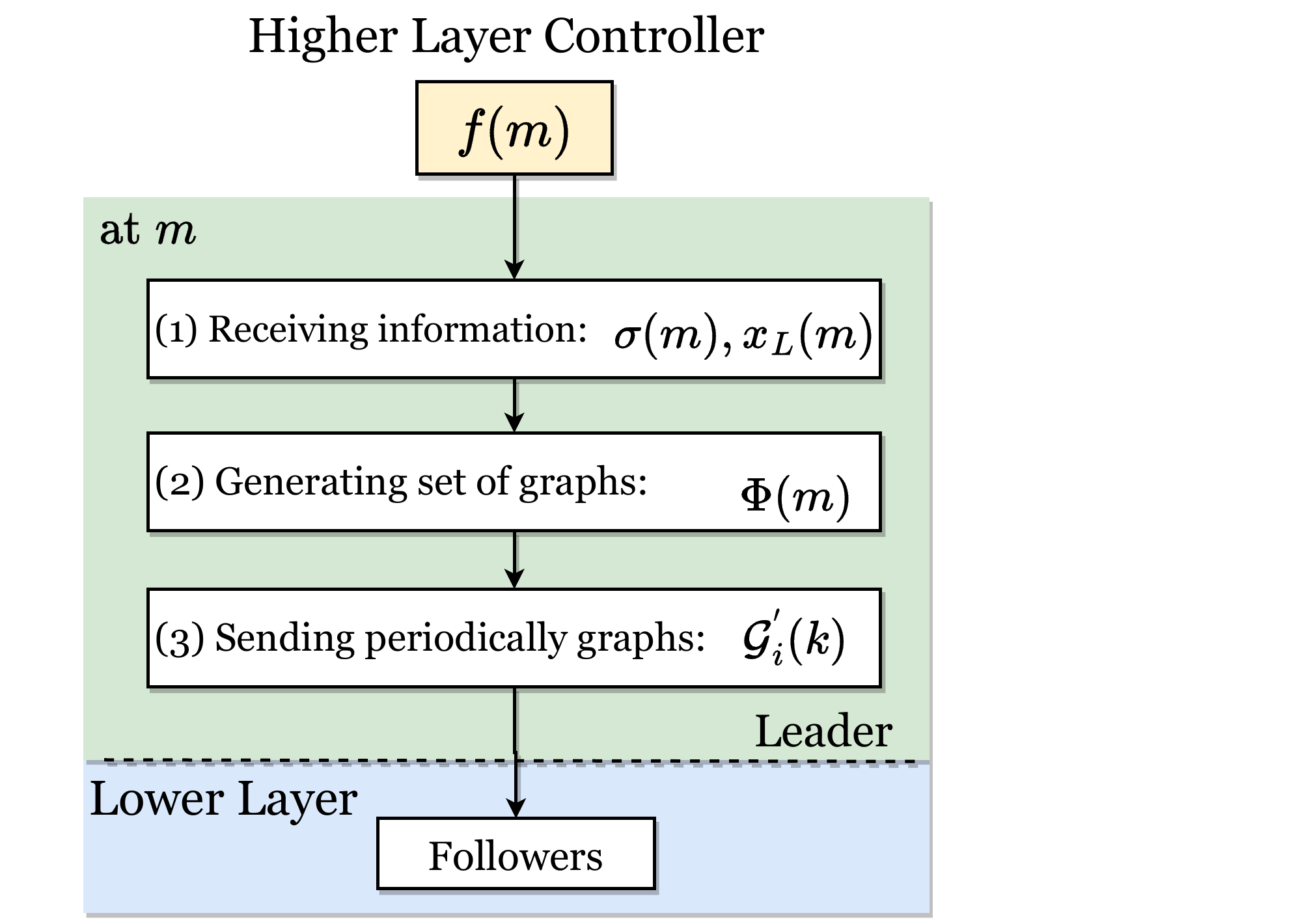}}
\caption{Flowchart depicting the proposed switching algorithm implemented in the leader of the considered TH-MAS}
\label{Fig:Flowchart_Proposition_Leader}
\end{figure}

Equation \eqref{eq:equality_constraints} constraints the number of controlled followers, yet we still have the freedom to select at each sampling time ($kT_s$) which followers are active (controlled). Motivated by this degree of freedom, we design a switching algorithm to determine which followers are controlled while \eqref{eq:equality_constraints} and \eqref{eq:piecewise_constraints} are met. This algorithm is implemented in the leader agent, which as described in Figure~\ref{Fig:Flowchart_Proposition_Leader} has three main tasks: receive information from the higher layer, generate a set of graphs and periodically send the active graph to the followers. 

Under requirements from \eqref{eq:equality_constraints} and \eqref{eq:piecewise_constraints}, we design a switching graph algorithm that periodically loops within a finite set of sorted communication graphs, i.e. 
\begin{equation}
    \label{eq:set_of_graphs_time_varying}
    \Phi(m) =  \{{\mathcal{G}^{'}_{0}}(k), \ldots,{\mathcal{G}^{'}_{p(m)-1}}(k) \},
\end{equation}
with 
\begin{equation}
    \label{eq:uniqueness_1}
    {\mathcal{G}^{'}_i(k)} \neq {\mathcal{G}^{'}_j(k)}.
\end{equation}
Note that the condition \eqref{eq:uniqueness_1} is true since
\begin{equation}
    \label{eq:uniqueness_2}
    {\mathcal{V}}^{'}_i  (k) \neq {\mathcal{V}}^{'}_j (k) \; \text{and} \; | {\mathcal{V}}^{'}_i  (k)| = |{\mathcal{V}}^{'}_j (k)|  
\end{equation}
for all $ i,j \in \{0,\ldots, p(m)-1\}$,  where
\begin{equation}
    \label{eq:number_of_combinations}
        p(m)= \frac{(N)!}{(N-\sigma(m))!\sigma(m)!} \in \mathbb{N}_+.
\end{equation}

\begin{figure}[htbp]
\centerline{\includegraphics[width=\columnwidth]{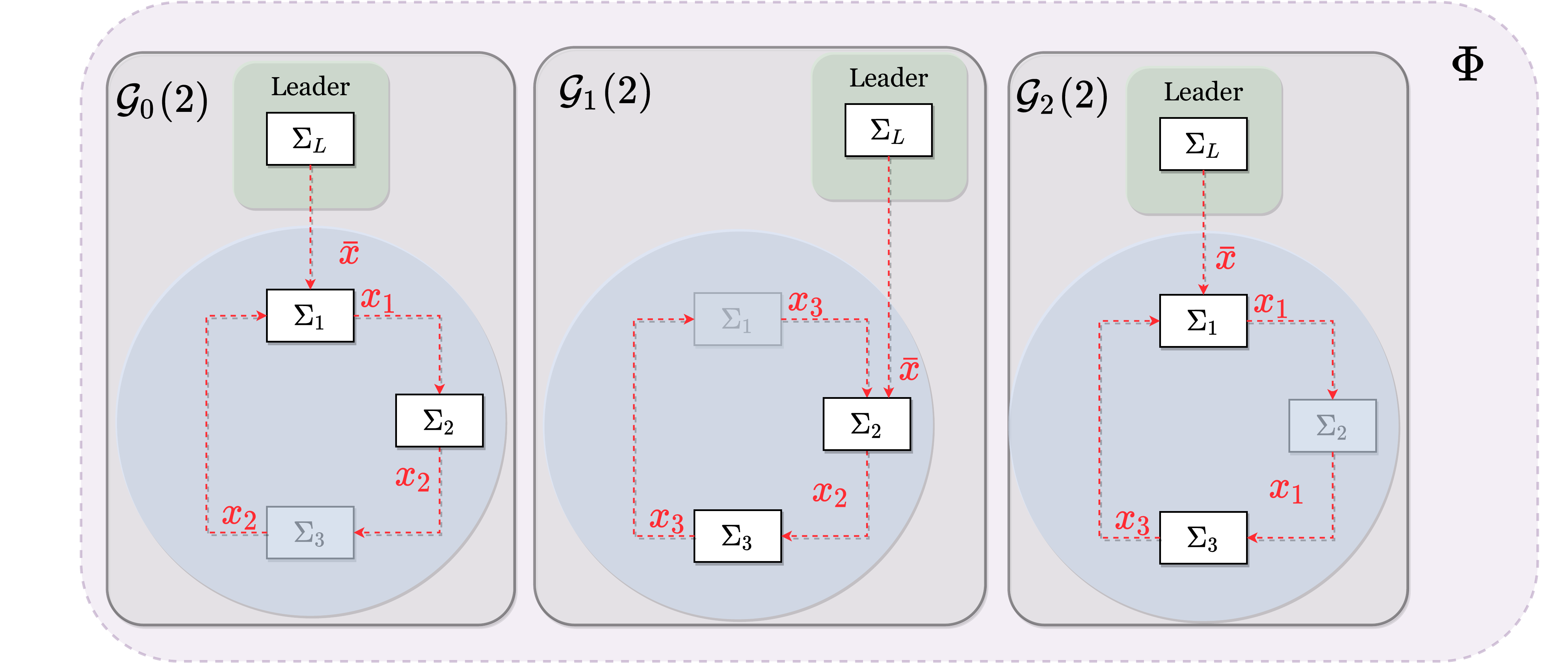}}
\caption{Illustration of the set $\Phi$ for a system with a number of agents $N=3$ and constrained to activate only 2 followers.}
\label{Fig:Switching_Network_Topology}
\end{figure}

Notice that \eqref{eq:number_of_combinations} shows how the number of combinations of selecting a specific number of followers from the entire system determines the cardinality of the set $\Phi(m)$. Equations \eqref{eq:number_of_combinations} and \eqref{eq:uniqueness_2} ensure the uniqueness of all the graphs within $\Phi (m)$. As depicted in Figure~\ref{Fig:Switching_Network_Topology}, the edges of nodes $( {\mathcal{E}}^{'}_i  (k))$ are linked such that the communication network topology is always the same. In this case, all $ {\mathcal{E}}^{'}_i  (k)$ have a ring topology. 

The proposed switching algorithm defines a time-varying graph, i.e. 

\begin{equation}
\label{eq:switching_piece_weise_d}
\boldsymbol{\mathcal{G}}(k) = \begin{cases} {\mathcal{G}^{'}_{0}(k)} \; \in \Phi (m), & \text { if } \sigma(m) \neq \sigma(m-1) \, \\ & \;\, \text{and} \;\;  k\,\text{mod} \, m=0,\\ {\mathcal{G}^{'}_{i(k)}(k)} \;  \in \Phi (m) & \text { else if } k\,\text{mod} \, m\neq 0\end{cases}
\end{equation}
where
\begin{equation}
\label{eq:index_for_switching}
{i}(k) = \begin{cases} i(k-1)+1  & \text { if }   i(k-1) < p(m), \\ 0 & \text { if } i(k-1) = p(m) \end{cases}
\end{equation}
with $i(k) \in \{0,1, \ldots, p(m)-1\}$.

\subsection{Consensus of TH-MAS: Lower Layer}
\label{sec:CONSENSUS_PROBLEM}
Consider a state feedback controller, such that, for the TH-MAS defined by \eqref{eq:follower_dynamics} and \eqref{eq:leader_dynamics}, the control inputs ($u_i$) are calculated as 
\begin{equation}
    \label{eq:consensus_protocol}
    C_{i}: u_{i}(k) = -k_{fb} \sum^{N+1}_{j=1, j\neq i } a_{i j}\left(x_{i}(k)-x_{j}(k)\right), \; 
\end{equation}
for all $i,j=\{1,\ldots,N+1\}$, where $k_{fb} \in \mathbb{R}$ is the feedback gain and $a_{ij}$ are the coefficients of the adjacency matrix $\mathcal{A}$.
Then, by implementing the switching algorithm \eqref{eq:switching_piece_weise_d} and state feedback controller \eqref{eq:consensus_protocol}, also called consensus protocol, in the above mentioned TH-MAS, the resulting closed-loop system is expressed by 
\begin{equation}
    \label{eq:overall_system_fedback_controller_discrete_time_varying_4.2}
     \bar{\boldsymbol{\Sigma}}_{d}:= \;\; {\boldsymbol{x}}(k+1) =  \boldsymbol{\mathcal{P}} (k) \cdot \boldsymbol{x}(k),
\end{equation}  
where  ${\boldsymbol{x}}(k) = [{x}_{1}, \ldots, {x}_{N+1} ]^{\top} \in \mathbb{R}^{N+1}$ is a vector containing the state of all the agents, and for all $k \in [(m-1)M+1,mM)$, i.e. for $q_L(k)=0$,
\begin{equation}
    \label{eq:defintion_of_time_varying_p_42}
    \boldsymbol{\mathcal{P}} (k) = \mathbf{I}_{N+1} -w \cdot k_{fb} \cdot {\mathcal{L}}(k) 
\end{equation}
 is a primitive row-stochastic matrix with ${\mathcal{L}}(k)={f_{\mathcal{L}}} {\left( \boldsymbol{\mathcal{G}}(k)  \right)},$ as in \eqref{eq:switching_piece_weise_d} and $k_{fb}$ as in \eqref{eq:consensus_protocol}. Note that $ a_{i j} = 0$ if the agents are not active; hence, the Laplacian matrix also describes the behaviour represented by $B(k)$.  In this paper, we focus on solving the consensus problem for a TH-MAS represented by \eqref{eq:overall_system_fedback_controller_discrete_time_varying_4.2}.
 
The consensus problem stands for the challenge of steering all the followers to a common state, i.e. 
\begin{equation}
    \label{eq:consensus_definition_for system}
    \lim_{k\rightarrow\infty} {\boldsymbol{x}}(k) =  \mathbf{1}_{N+1} \bar{x}
\end{equation}
where ${\boldsymbol{x}}(k) = [{x}_{1}, \ldots, {x}_{N+1} ]^{\top} \in \mathbb{R}^{N+1}$ is a vector containing the state of all the agents and $\bar{x} \in \mathbb{R}$ is the consensus value. 
It is important to remark that $\bar{x}$ is determined by the Laplacian matrix and the initial states of the agents in the system as the following lemma explains. 
\begin{lem}
\label{lmm:invarint_property_consensus_dynamics}
[Lemma 3.1,\citep{Lunze2019}]  Consider system $\bar{\boldsymbol{\Sigma}}_{d}$ as in \eqref{eq:overall_system_fedback_controller_discrete_time_varying_4.2} with ${\mathcal{L}}(k) = \mathcal{L}$, for all $k \in \mathbb{N}$ where $\mathcal{L}$ has a spanning tree. The system is invariant under movement, i.e. 
$$
y=\boldsymbol{w}^{\top} \boldsymbol{x}(k), \quad \forall k \in \mathbb{N},
$$
where $\boldsymbol{w}^{\top} \geq 0^{\top}$ is a left eigenvector of the Laplacian matrix $\mathcal{L}$ for the vanishing eigenvalue $\lambda_1=0$ and $y \in \mathbb{R}$ is a scalar number that represents the consensus value. 
\end{lem}

\begin{rem}
    \label{rmk:important}
    In Lemma~\ref{lmm:invarint_property_consensus_dynamics}, $\boldsymbol{w}^{\top}$ can also be defined as the left eigenvector associated to $\lambda =1$ of the corresponding time-invariant primitive matrix $\mathcal{P}$ calculated as in \eqref{eq:defintion_of_time_varying_p_42} with ${\mathcal{L}}(k) = \mathcal{L}$, for all $k \in \mathbb{N}$. This is verified by modifying system \eqref{eq:overall_system_fedback_controller_discrete_time_varying_4.2} such that 
    \begin{equation}
        \label{eq:proof_w_c}
        \boldsymbol{w}^{\top} \boldsymbol{x}(k +1) =  \boldsymbol{w}^{\top} \mathcal{P}   \boldsymbol{x}(k)  =  \boldsymbol{w}^{\top}   \boldsymbol{x}(k),  
    \end{equation}
    where the last equality follows due to the property of the eigenvector $\boldsymbol{w}^{\top}$ corresponding to a primitive stochastic matrix $\mathcal{P}$, proving that the system \eqref{eq:overall_system_fedback_controller_discrete_time_varying_4.2} is invariant under movement.
\end{rem}

Equation \eqref{eq:consensus_protocol} considers a communication network with a constant topology. Yet, in \citep{Ren2005}, it is shown that with static feedback gain, as in \eqref{eq:consensus_protocol}, one can steer all the agents with a single integrator dynamics to achieve asymptotically consensus, even when the communication network changes its topology. The following lemma was crucial to formalize the above statement.

\begin{lem}
\label{lmm:union_graphs_convergence}
[Lemma 3.9,\citep{Ren2005}] If the union of a set of directed graphs $\left\{\mathcal{G}_{{1}}, \mathcal{G}_{{2}}, \ldots, \mathcal{G}_{{m}}\right\} \subset \mathcal{G}$, where $m\in\mathbb{N_+}$, has a spanning tree, then the matrix product $P_{{m}} \ldots P_{{2}} P_{{1}}$ is primitive row-stochastic, where $P_{{i}}$ is a stochastic matrix corresponding to each directed graph $\mathcal{G}_{{i}}$ as in \eqref{eq:defintion_of_time_varying_p_42}.
\end{lem}

From \eqref{eq:switching_piece_weise_d}, it is observed that he union of all the graphs within $\Phi(m)$ results in a new graph, i.e.
\begin{equation}
    \label{eq:union_switching_algorithm}
    \boldsymbol{\Bar{\mathcal{G}}} =  \bigcup_{i=0}^{p-1} {{\mathcal{G}}^{'}_{i}(k) }, \; \;  {{{\mathcal{G}}^{'}_{i}(k) } \in \Phi(m)},
\end{equation}
where $\boldsymbol{\Bar{\mathcal{G}}} = (\boldsymbol{\Bar{\mathcal{V}}}, \boldsymbol{\Bar{\mathcal{E}}})$. Since the sorted set of graphs $\Phi$ contains all the possible combinations of choosing certain number of nodes from $\bar{\mathcal{V}}$, then  ${\boldsymbol{\Bar{\mathcal{V}}}}' = \bar{\mathcal{V}}$. Moreover, $\boldsymbol{\Bar{\mathcal{E}}}$ describes a communication network where the followers are fully connected because ${\boldsymbol{\Bar{\mathcal{V}}}}' = \bar{\mathcal{V}}$ and ${{{\mathcal{G}}^{'}_{i}(k) }}$ have a spanning three.  Hence, $\boldsymbol{\Bar{\mathcal{G}}}$ is a graph with a spanning tree that covers all the agents within ${\mathcal{G}}$. 

\begin{figure}[htbp]
\centerline{\includegraphics[width=5cm]{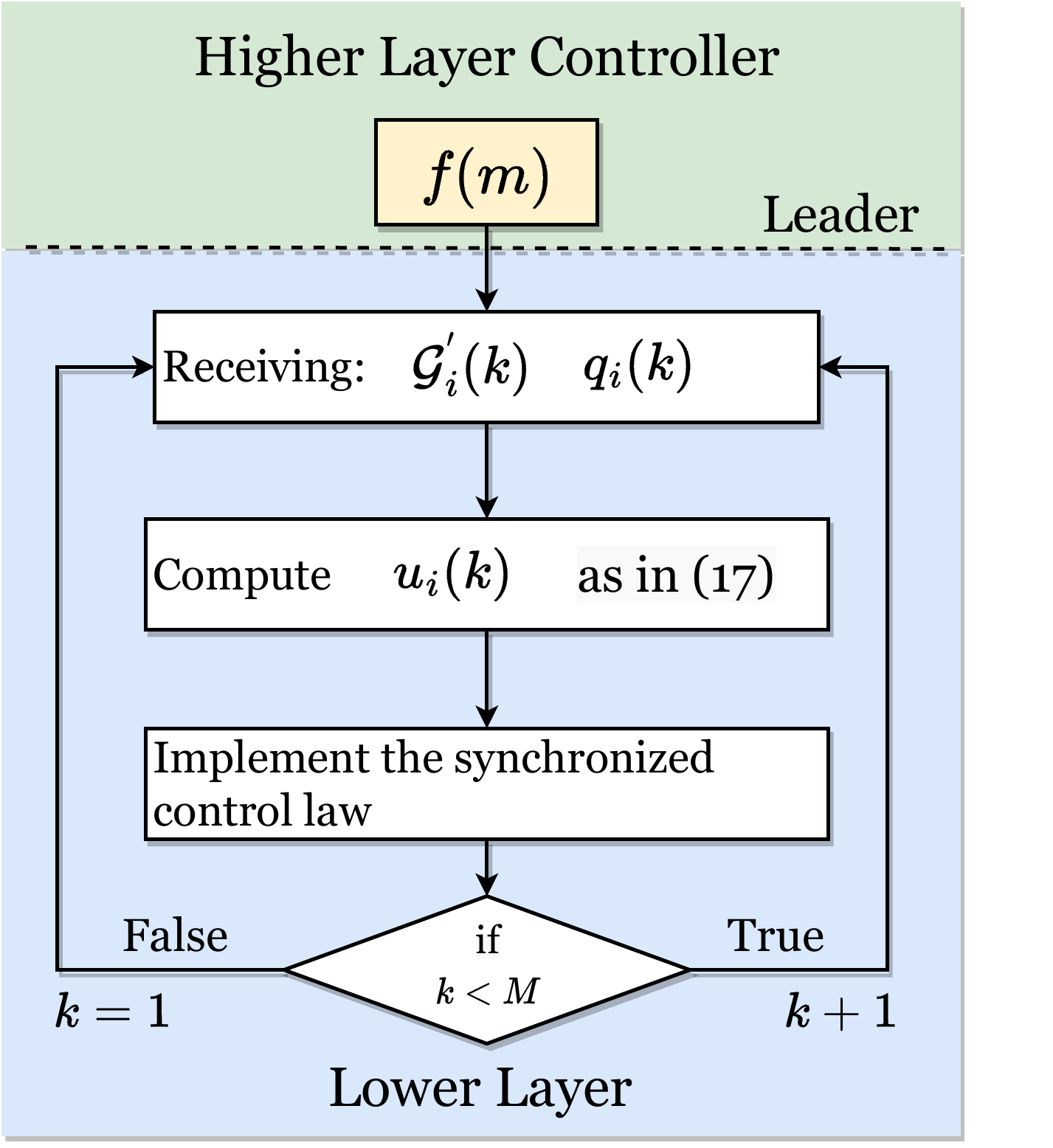}}
\caption{Flowchart depicting the proposed control algorithm implemented in the followers considered TH-MAS}
\label{Fig:Flowchart_Proposition}
\end{figure}

The proof of \citep{Ren2005} does not include the case when the set of active agents varies with time. Yet, the notions explained in  \citep{Ren2005} are used to implement a solution for the considered TH-MAS. In this case, the control law as in \eqref{eq:consensus_protocol} subject to the changing graphs from the switching algorithm is applied to all the active followers, see Figure~\ref{Fig:Flowchart_Proposition}. Note that for proving the asymptotic convergence of the  proposed solution the following assumption is made.
\begin{assum}
    \label{assum:change_}
    Consider that $x_L(k)$ does not change, i.e.
    $$ q_L(k) = 0, \; \forall k \in \mathbb{N}_+ $$  and $x_L(0) = \bar{x}$.
\end{assum}


\begin{thm}
\label{prop:convergence}
Consider the TH-MAS as in \eqref{eq:follower_dynamics}-\eqref{eq:leader_dynamics} with the constraints \eqref{eq:equality_constraints}, a switching graph $\boldsymbol{\mathcal{G}}(k)$ as in \eqref{eq:switching_piece_weise_d}, and a control input $(u_i(k))$ as in \eqref{eq:consensus_protocol}, such that 
\begin{equation}
    \label{eq:consensus_protocol_tunning}
     0<k_{fb}<\frac{1}{\max(d_i)w},
\end{equation}
where $d_i \in \mathbb{N}$ are the elements in the diagonal of the degree matrix $(\mathcal{D})$. Suppose that Assumption~\ref{assum:change_} holds. 
Then, all the followers reach a consensus state, i.e.
\begin{equation}
    \label{eq:consensus_definition_for_system_2}
    \lim_{k\rightarrow\infty} {\boldsymbol{x}}(k) = \bar{x} \mathbf{1}_{N+1}
\end{equation}
where $\bar{x} = x_L(0)$ is the consensus state.
\end{thm}

\begin{pf}
\label{proof:mine_only_one}
Consider a system with the agent dynamics as \eqref{eq:follower_dynamics} and \eqref{eq:leader_dynamics}, a controller as \eqref{eq:consensus_protocol} and the switching algorithm from  \eqref{eq:switching_piece_weise_d}. Then a closed-loop system dynamics is defined as in \eqref{eq:overall_system_fedback_controller_discrete_time_varying_4.2}.

Due to \eqref{eq:consensus_protocol_tunning}, for all $k\in \mathbb{N}$, $ {\mathcal{G}^{'}_{i}(k)} $ has a spanning tree. Consequently, \eqref{eq:defintion_of_time_varying_p_42} is primitive row-stochastic, meaning that $\boldsymbol{\mathcal{P}} (k)$ is nonnegative and has positive diagonal elements. That can easily observed if \eqref{eq:defintion_of_time_varying_p_42} is rewritten as
\begin{equation}
    \label{eq:primitive_matrix_defnition_1}
    \boldsymbol{\mathcal{P}} (k) = \mathbf{I}_{N+1} - w k_{fb}  {\mathcal{D}}(k) + w k_{fb}  {\mathcal{A}}(k).
\end{equation}

From \citep{wolfowitz1963}, it is known that any primitive row-stochastic matrix ($\mathcal{P} \in \mathbb{R}^{n\times n}$ )  has an invariant path to the steady-state, i.e.
\begin{equation}
    \label{eq:sia_convegence_condition}
    \lim_{j\rightarrow\infty}{\left(\mathcal{P}\right)}^{j} =  \mathbf{1}_{n} c^{\top},
\end{equation} 
where $n$ is the dimension of matrix $\mathcal{P}$ and $c \in \mathbb{R}^{n}$ is a column vector representing the left eigenvector of $\mathcal{P}$ associated with the eigenvalue $\lambda = 1$. Note that we call an invariant path the term in the right-hand side of \eqref{eq:sia_convegence_condition}. 

Given \eqref{eq:union_switching_algorithm}, it is known that union of all the graphs ($\boldsymbol{\Bar{\mathcal{G}}}$) is invariant with respect to $\sigma(m)$, covers all the agents and has a spanning tree. Hence, using Lemma~\ref{lmm:union_graphs_convergence} and the property from \eqref{eq:sia_convegence_condition}, we can infer that \eqref{eq:overall_system_fedback_controller_discrete_time_varying_4.2} will have an invariant path as in \eqref{eq:sia_convegence_condition} and will reach consensus. The previous statement is verified in three steps: 

\begin{enumerate}
    \item[(i)] Given $\boldsymbol{x}(0)$, consider the solution of \eqref{eq:overall_system_fedback_controller_discrete_time_varying_4.2}  as the infinite sequence of $\boldsymbol{x}(k)  \in  \mathbb{R}^{N+1}$, i.e.
    \begin{equation}
    \label{eq:sequence_system_}
       \boldsymbol{x}(k+1) =  \boldsymbol{\mathcal{P}} (k) \boldsymbol{x}(k) \quad  \text{with} \;\, k \in \mathbb{N}.
    \end{equation}
    
    \item[(ii)] Since $\boldsymbol{\mathcal{P}} (k)$ has a periodic behaviour, due to \eqref{eq:switching_piece_weise_d}, decompose $\boldsymbol{x}(k)$ in a finite number of subsequences $b^{i}_{n}$, such that  
    \begin{equation}
        \label{eq:union_of_subsequence}
         \boldsymbol{x}(k) = \bigcup_{i=0}^{p-2} b^{i}_{n}
    \end{equation}
    with 
    \begin{equation}
        \label{eq:subsequence_definition}
        \begin{aligned}
            b^{0}_{n} &= \{\boldsymbol{x}(0),  \boldsymbol{x} \left( p-1\right),  \boldsymbol{x}\left( 2p-2\right), \ldots\} \\ 
            b^{1}_{n} &= \{\boldsymbol{x}(1),  \boldsymbol{x}\left( p\right),  \boldsymbol{x}\left( 2p-1\right), \ldots\} \\
            b^{2}_{n} &= \{\boldsymbol{x}(2),  \boldsymbol{x}\left( p+1\right),  \boldsymbol{x}\left( 2p\right), \ldots\} \\
            \vdots  \\
            b^{p-2}_{n} &= \{\boldsymbol{x}(p-2),  \boldsymbol{x}\left( 2p-3\right),  \boldsymbol{x}\left( 3p-5 \right), \ldots\} \\
        \end{aligned}
    \end{equation}
    where $n \in \mathbb{N}$ are the sort indices of the subsequences and $p$ is the number of graphs in the set $\Phi$ as in \eqref{eq:number_of_combinations}. Notice that the $n^{\text{th}}$ element of the subsequences $b^{i}_{n}$ corresponds to the  $k^{\text{th}}$ element of the sequence $\boldsymbol{x}(k)$, such that $k = (p-1)n + i$.\\
    
    \item[(iii)] Rearrange the subsequences as 
    \begin{equation}
    \label{eq:subsequence_formulation_as_function_of_b}
    {b^{i}_{0}} =  \boldsymbol{x}(i), \quad {b^{i}_{n+1}} = \overline{\mathcal{P}}_{i} \, {b^{i}_{n}}, \quad  \forall i={0, \ldots, p-2},
    \end{equation}
    with $$ \overline{\mathcal{P}}_{i} = {\mathcal{P}}_{i-1} \times {\mathcal{P}}_{i-2} \times \ldots \times {\mathcal{P}}_{i}, $$
    where $\overline{\mathcal{P}}_{i}$ is the product of the matrices ${\mathcal{P}}_{i}$ shifted in different orders depending on $i$. Consequently, $\overline{\mathcal{P}}_{i}$ are primitive row-stochastic matrices, since $$\overline{\mathcal{P}}_{i} = \mathbf{I}_{N+1} -w \cdot k_{fb} \cdot {f_{\mathcal{L}}} {\left( \boldsymbol{\Bar{\mathcal{G}}}(k)  \right)}, \;  \forall i \in\{0, \ldots, p-2\} .$$ 
    Given \eqref{eq:union_switching_algorithm} and  Lemma~\ref{lmm:union_graphs_convergence}, all the subsequences $b^{i}_{n}$ have the same invariant path and converge to a value determined by their initial conditions, i.e. $b^i_0 = \boldsymbol{x}(i) $, such that
    \begin{equation}
    \label{eq:limit_subsequences}
    \resizebox{.9\hsize}{!}{$ \begin{aligned}
            \lim _{n \rightarrow \infty} b_{n}^{0}         &= \lim _{n \rightarrow \infty} {\underbrace{\left(\mathcal{P}_{(p-1)} \times \mathcal{P}_{(p-2)} \times \ldots \times \mathcal{P}_{0}\right)}_{\overline{\mathcal{P}}_{0}}}^{n} \cdot \boldsymbol{x}(0) \\
            \lim_{n \rightarrow \infty} {b^{1}_{n}}        &= \lim _{n \rightarrow \infty} {\underbrace{\left(\mathcal{P}_{(0)} \times \mathcal{P}_{(p-1)} \times \ldots \times \mathcal{P}_{1}\right)}_{\overline{\mathcal{P}}_{1}}}^{n} \cdot \boldsymbol{x}(1) \\
            \lim_{n \rightarrow \infty} {b^{2}_{n}}        &= \lim _{n \rightarrow \infty} {\underbrace{\left(\mathcal{P}_{(1)} \times \mathcal{P}_{(0)} \times \ldots \times \mathcal{P}_{2}\right)}_{\overline{\mathcal{P}}_{2}}}^{n} \cdot \boldsymbol{x}(2) \\
            \vdots \\
            \lim_{n \rightarrow \infty} {b^{(p-2)}_{n}}    &= \lim _{n \rightarrow \infty} {\underbrace{\left(\mathcal{P}_{(p-3)} \times \mathcal{P}_{(p-4)} \times \ldots \times \mathcal{P}_{0}\right)}_{\overline{\mathcal{P}}_{(p-2)}}}^{n} \cdot\boldsymbol{x}(p-2),
        \end{aligned}
    $}    
    \end{equation}  
    since
    \begin{equation}
        \label{eq:same_path_equation}
        \lim _{n \rightarrow \infty} {\overline{\mathcal{P}}_{0}}^{n} \, = \, \ldots \,  =  \, \lim_{n \rightarrow \infty} {\overline{\mathcal{P}}^{n}_{(p-2)}} \, = \,  \mathbf{1}_{N+1} c^{\top}, 
    \end{equation} as stated in \eqref{eq:primitive_matrix_defnition_1}. Hence,
    \begin{equation}
        \label{eq:limit_subsquence_final}
        \lim _{n \rightarrow \infty} b_{n}^{i} = \lim _{n \rightarrow \infty} {\overline{\mathcal{P}}_{i}}^{n} \boldsymbol{x}(i) = \mathbf{1}_{N+1} c^{\top} \boldsymbol{x}(i)
    \end{equation}
    for all $i= {0,\ldots,p-2}.$
\end{enumerate}


Notice that based on \eqref{eq:same_path_equation}, all subsequences have the same path, i.e. $ \mathbf{1}_{N+1} c^{\top}.$ Recalling Lemma~\ref{lmm:invarint_property_consensus_dynamics}, Remark~\ref{rmk:convergence_time_ } and knowing that $ {f_{\mathcal{L}}} {\left( \boldsymbol{\Bar{\mathcal{G}}}(k)  \right)}$ has a spanning tree, the following equality holds
\begin{equation}
    \label{eq:equality_very_important}
    \lim _{n \rightarrow \infty} b_{n}^{i} = \mathbf{1}_{N+1} \boldsymbol{w}^{\top} \boldsymbol{x}(i), \quad \forall i= {0,\ldots,p-2}.
\end{equation}
The Laplacian matrix $\boldsymbol{\overline{{\mathcal{L}}}}$ that represents  $\boldsymbol{\Bar{\mathcal{G}}}$ connects all the followers to the leader. Then, as shown in \citep{Lunze2019}, it holds that the last row of the Laplacian matrix vanishes and the matrix can be rearranged as 
$$\boldsymbol{\overline{{\mathcal{L}}}} = \left(\begin{array}{c|c}
0 & \boldsymbol{0}_{N}^{\top} \\
\hline \boldsymbol{\tilde{{\mathcal{L}}}} & l
\end{array}\right).$$
Above, the first row and the first column have been separated from the remaining $N$ rows and columns, $l^{\top} = [0\; 0\;  \ldots \; 1] \in \mathbb{N}^{N}$ and  $ \boldsymbol{\tilde{{\mathcal{L}}}}$ is the Laplacian matrix corresponding to the communication network that contains all the followers. 

For all matrices as $\boldsymbol{\overline{{\mathcal{L}}}}$, as it is proven in \citep{Lunze2019}, the normalised vector is defined as $$\boldsymbol{w}^{\top} = [ 0\;\; 0 \;\; \ldots \;\; 1] \in \mathbb{R}^{N+1}.$$ Hence, the following equality hold,
\begin{equation}
    \label{eq:very_important_equalities}
    \boldsymbol{w}^{\top} \boldsymbol{x}(i) = x_L(k) \; \forall i= {0,\ldots,p-2}.
\end{equation}
Then, for all $i=0,...,p-2$,
\begin{equation}
        \label{eq:limit_subsquence_final_2}
        \lim _{n \rightarrow \infty} b_{n}^{i}  = \mathbf{1}_{N+1}\boldsymbol{w}^{\top} \boldsymbol{x}(i) =  \mathbf{1}_{N+1} \bar{x}.
\end{equation}
Since the finite number of subsequences $b^{i}_{n}$ whose union is equal to the original sequence $\boldsymbol{x}(k)$ converge to the same limit, then also the sequence $\boldsymbol{x}(k)$ will converge to the same limit (see, for example, Theorem~3.5.7 from~\citep{Aguilar2022}). Consequently,  as in Lemma~\ref{lmm:invarint_property_consensus_dynamics}, the states of system \eqref{eq:overall_system_fedback_controller_discrete_time_varying_4.2} converge to a value that is considered the consensus value, i.e. $\bar{x}=x_L(k)$. This complete the proof. \qed
\end{pf}
\begin{rem}
\label{rmk:convergence_time_ }
Theorem~\ref{prop:convergence} only proves the asymptotic convergence of the followers states to a consensus value defined by the leader state for a constant $ x_L(k)$, i.e., for $ q_L(k)=0$ for all $k\in \mathbb{N}$. In a realistic TH-MAS system, however, $x_L(k)$ will be a piece-wise constant signal rather than a constant one due to the update from $u_L(k)$ at $k=mM$. In this case, as it will be illustrated in the simulation examples, practical consensus will be achieved if the time difference between the two layers is sufficiently large, i.e., to allow switching through all of the graphs in the set $\Phi (m)$ multiple times. Additionally, it is worth noting that $k_{fb}$ and the topology of $\boldsymbol{\Bar{\mathcal{G}}}$, which must have a spanning tree, determine the transient dynamics under the consensus control law.
\end{rem}

\section{Illustrative examples}
\label{sec:Case_of_Study}

\subsection{{Modular Multilevel Converter Example}}
\label{subsec:Modular_Multilevel_Converter_Example}
Balancing Capacitor Voltage (BAC) is a common control problem in Modular Multilevel Converters (MMCs). In this case, all the followers (the modules) need to store a certain amount of voltage $(v_C)$  in their capacitors such that the total stored voltage in the capacitors $(v^{\Sigma}_C)$ is equal to a voltage reference $({v^{{\Sigma}^*}_{C}})$ coming from a higher layer. The followers dynamics is such that,
\begin{equation}
    \label{eq:modules_dynamics_}
    \Sigma_{i}: {x}_{i}(k+1) =  x_{i}(k) -  w  q_i(k) u_{i}(k), 
\end{equation}
for all $i=\{1,2, \ldots, N\}$, where  $k \in \mathbb{N}$, $T_s$ is the sampling period, $x_i \in \mathbb{R}$ is the state variable describing the $i^{th}$ module capacitor voltage $(v^{i}_{C})$, $u_i \in  \mathbb{Z}_{[-1,1]}$ is the control input describing the switching signal $(n^r_i)$ provided to the modules and $w = \frac{\imath N T_s}{C} \in \mathbb{R}$, where $N$ is the number of modules in each arm, $C = 0.4 m$F is the module  capacitance, $\imath =80$A  is the common exogenous input representing the arm current.

In an MMC, all modules (follower agents) must have a uniform distribution of energy, which means that, at any sampling time, the same voltage must be stored in their capacitors. This is why a consensus on the capacitor voltages is needed.  Hence, the leader state is defined as
\begin{equation}
    \label{eq:leader_dynamics_mmc}
    x_L(k) =  \frac{{v^{{\Sigma}^*}_{C}}(k)}{N}, \,\forall k=mM,
\end{equation}
where $N$ is the number of existing modules. The switching algorithm is also needed due to hard constraints regarding of the number of modules that can be active at each sampling time. 
In this case, the communication topology in all the graphs is a ring network.

\begin{figure}[htbp]
\centerline{\includegraphics[width=\columnwidth]{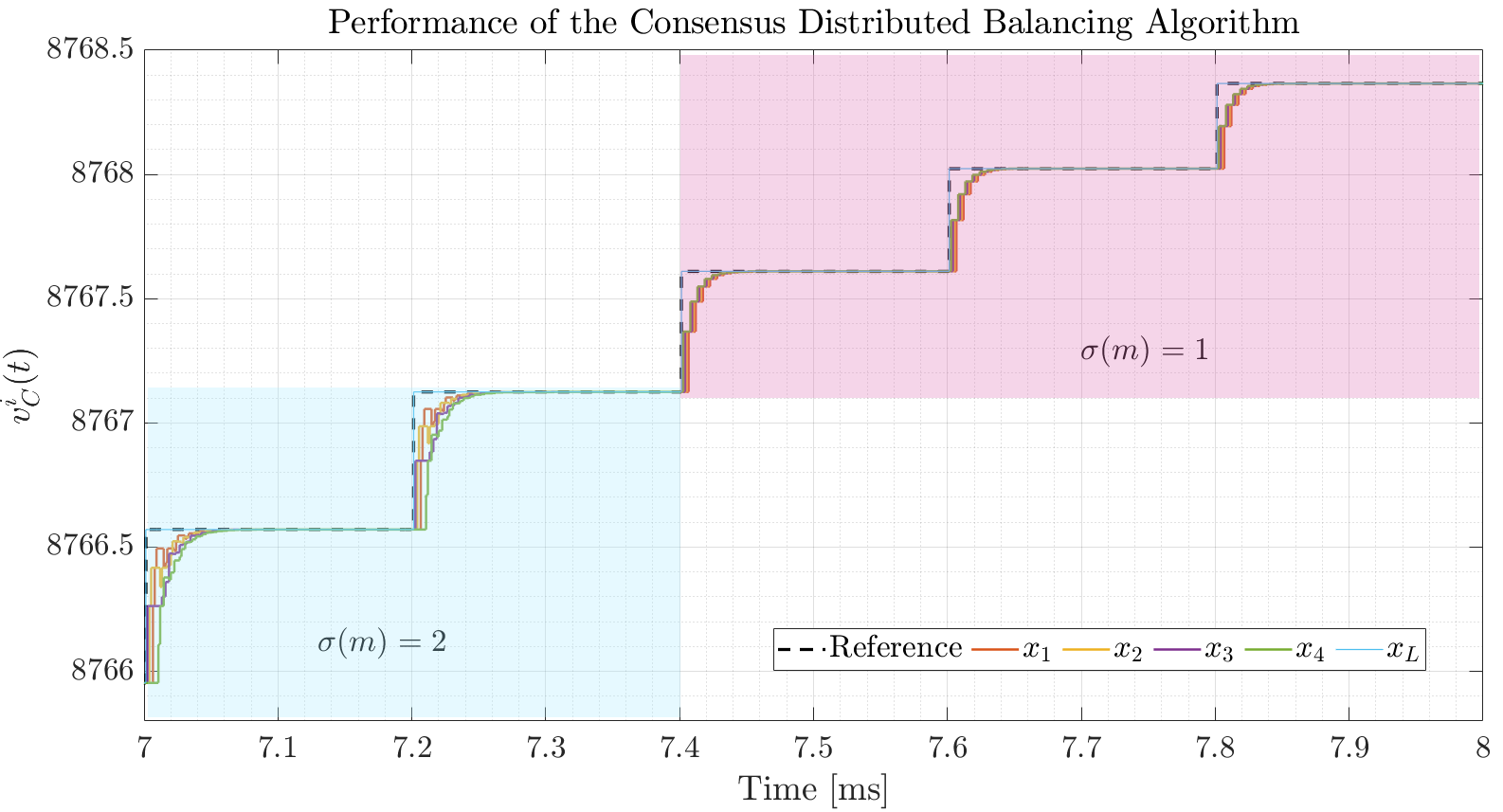}}
\caption{Capacitor voltages using the switching algorithm in \eqref{eq:index_for_switching} with continuous input.}
\label{Fig:Modules_Voltage_Cont}
\end{figure}

Figure 6 shows how the followers can reach consensus with the leader even when, due to $\sigma(m)$ the set of active agents only can contain 2 and 1 follower, respectively. In this case, the leader's behaviour is piecewise constant and the dynamics of the lower layer is fast enough to facilitate practical asymptotic convergence. Yet, in a real MMC circuit, the control input $u_i$ is a discrete input that can have only 3 possible values, i.e. -1, 0 or 1; hence, we define the control law as in \eqref{eq:consensus_protocol} with $k_{fb}=0.2$ and project the switching as
\begin{equation}
\label{eq:casting_control_input}
n^r_{i}(k)= \begin{cases}1 & \text { if } u_i(k) >0,\\ -1 & \text { if } u_i(k) <0, \\ 0 & \text { if } u_i(k) =0.\end{cases}
\end{equation}

\begin{figure}[htbp]
\centerline{\includegraphics[width=\columnwidth]{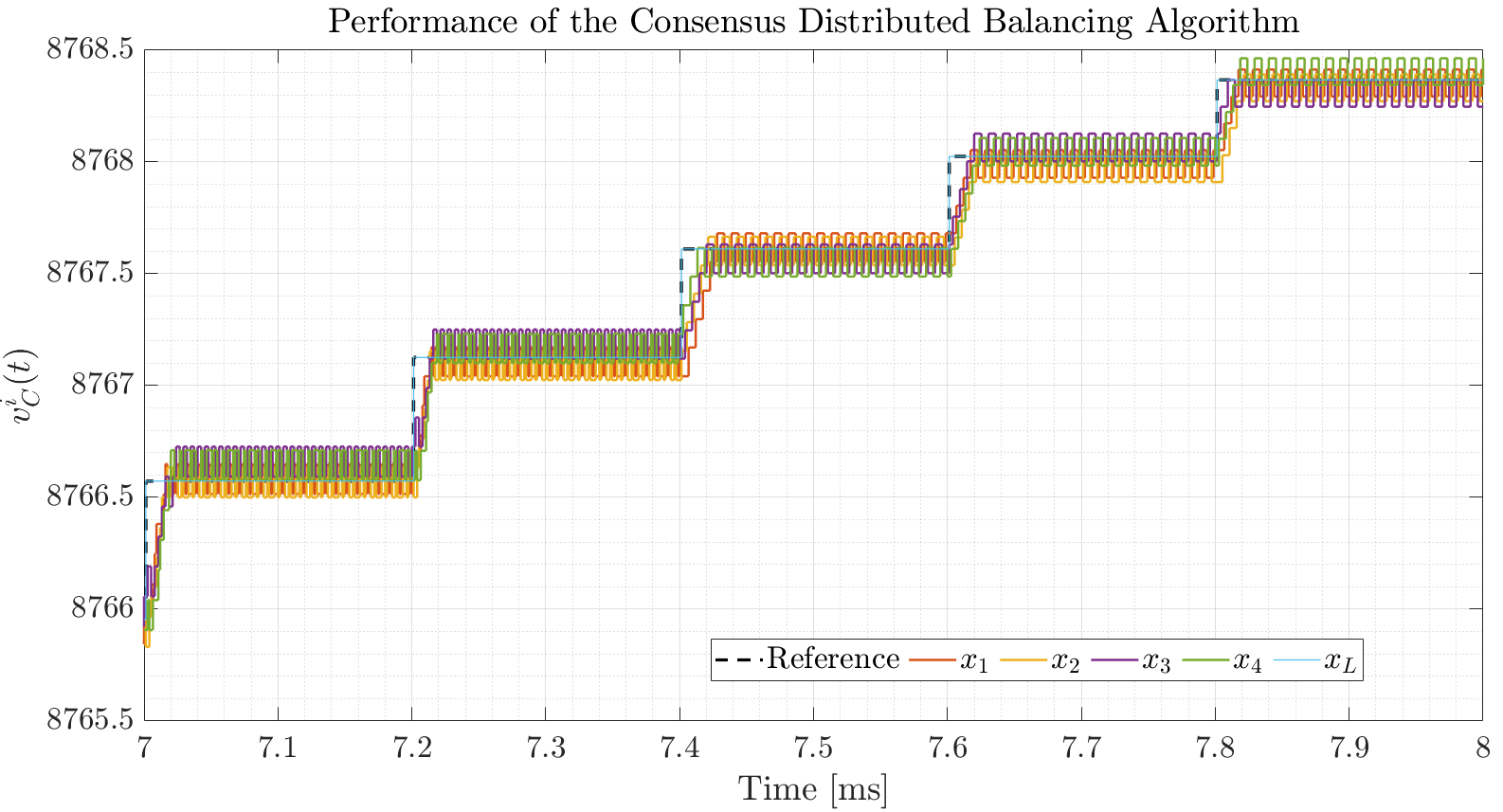}}
\caption{Capacitor voltages using switching algorithm in \eqref{eq:index_for_switching} and projected control input as in \eqref{eq:casting_control_input}.}
\label{Fig:CModules_Voltage}
\end{figure}

In Figure~\ref{Fig:CModules_Voltage},  we present the results when control input is projected as a switching signal as in \eqref{eq:casting_control_input}. Notice that a limit cycle around the desired value is reached. This result aligns with the conclusion presented in \citep{Morita2018} where a practical consensus for MAS with quantized inputs was defined. 
\begin{figure}[htbp]
\centerline{\includegraphics[width=\columnwidth]{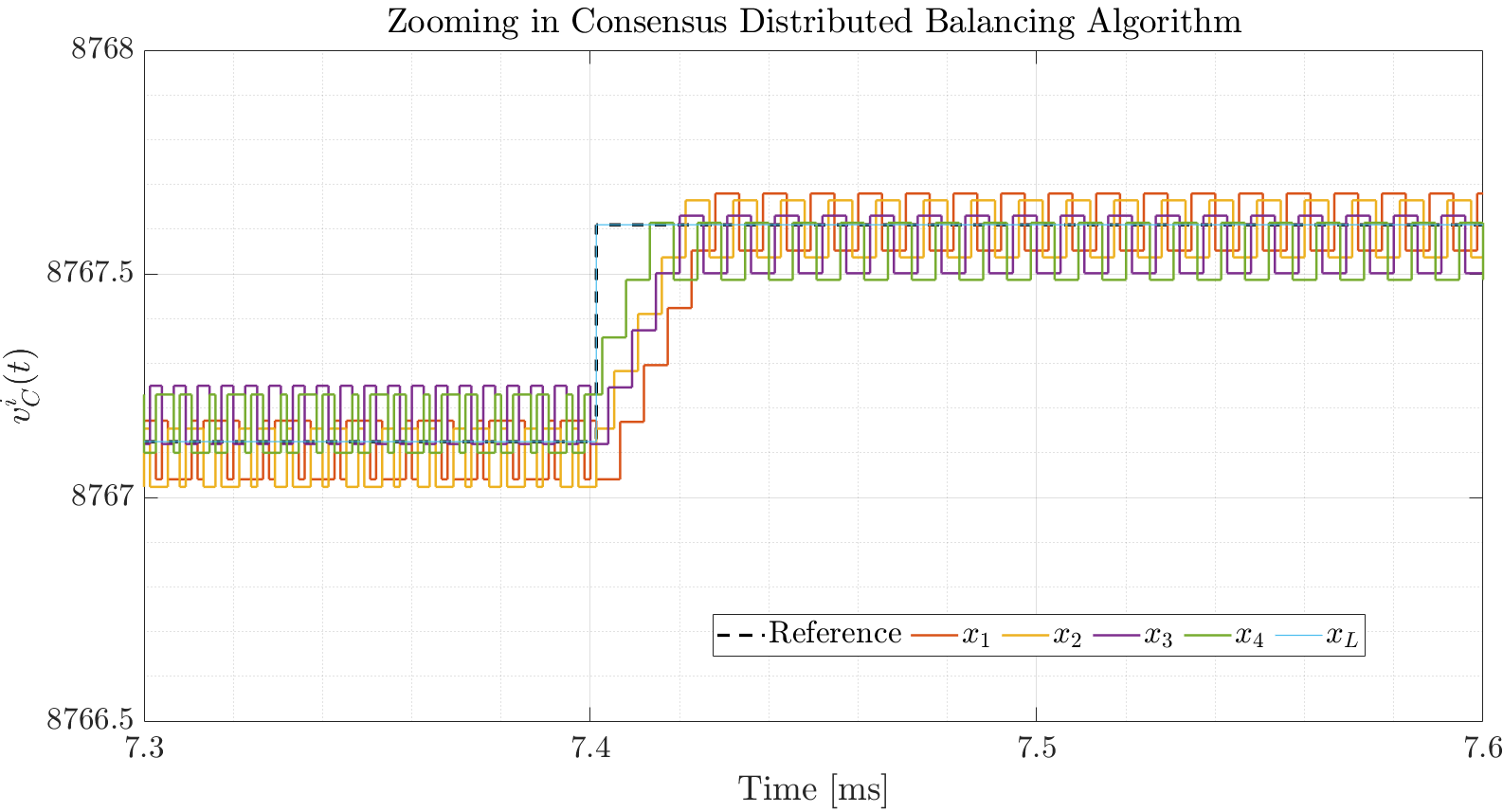}}
\caption{Zoom into a region of  Figure~\ref{Fig:CModules_Voltage} showing different limit cycles for different numbers of active agents}
\label{Fig:Modules_Voltage}
\end{figure}

In Figure~\ref{Fig:Modules_Voltage}  a zoom into a region of Figure~\ref{Fig:CModules_Voltage} is made. We observe that before $t=7.4$ms, two modules are activate per sampling time $k$ due to $\sigma(m)=2$. Meanwhile, after $t=7.4$ms, one module is activate per sampling time $k$ due to $\sigma(m+1)=1$. In this case, changing the number of allowed active modules $\sigma (m)$ did not affect the capacity of our method to converge to the desired value. 
\subsection{Water pumping system}
\label{subsec:Case_of_Study}
In water pumping systems, it is needed to supply a certain flow $(Q^*)$. In this case, the main challenge is to coordinate all the pumps to supply the desired flow while keeping the number of active pumps limited. The motivation for this constraint is to keep a proper balance among the lifespans of the pumps, where the working time of the pumps should be similar among all of them. This is why the switching algorithm is a suitable solution.

In this case, we consider the system presented in \citep{Kusumawardana2019}, where the flow ($Q = K*\omega$) is proportional to the speed rotation of the pump ($\omega$) and is expressed as 
\begin{equation}
    \label{eq:modules_dynamics_pumps}
    \Sigma_{i}: {x}_{i}(k+1) =  x_{i}(k) - w q_i(k)   u_{i}(k), 
\end{equation}
for all $i=\{1,2, \ldots, N\}$, where  $k \in \mathbb{Z_+}$, $T_s$ is the sampling period,  $x_i \in \mathbb{R}$ is the state variable describing the $i^{th}$ pump flow $(Q_i)$, $w = 0.2e^3 \frac{m^2}{\text{sA}} $  is a constant defined by pump characteristics and $u_i \in  \mathbb{R}$ is the control input describing the pump armature current $(Ia_i)$ .
\begin{figure}[htbp]
\centerline{\includegraphics[width=\columnwidth]{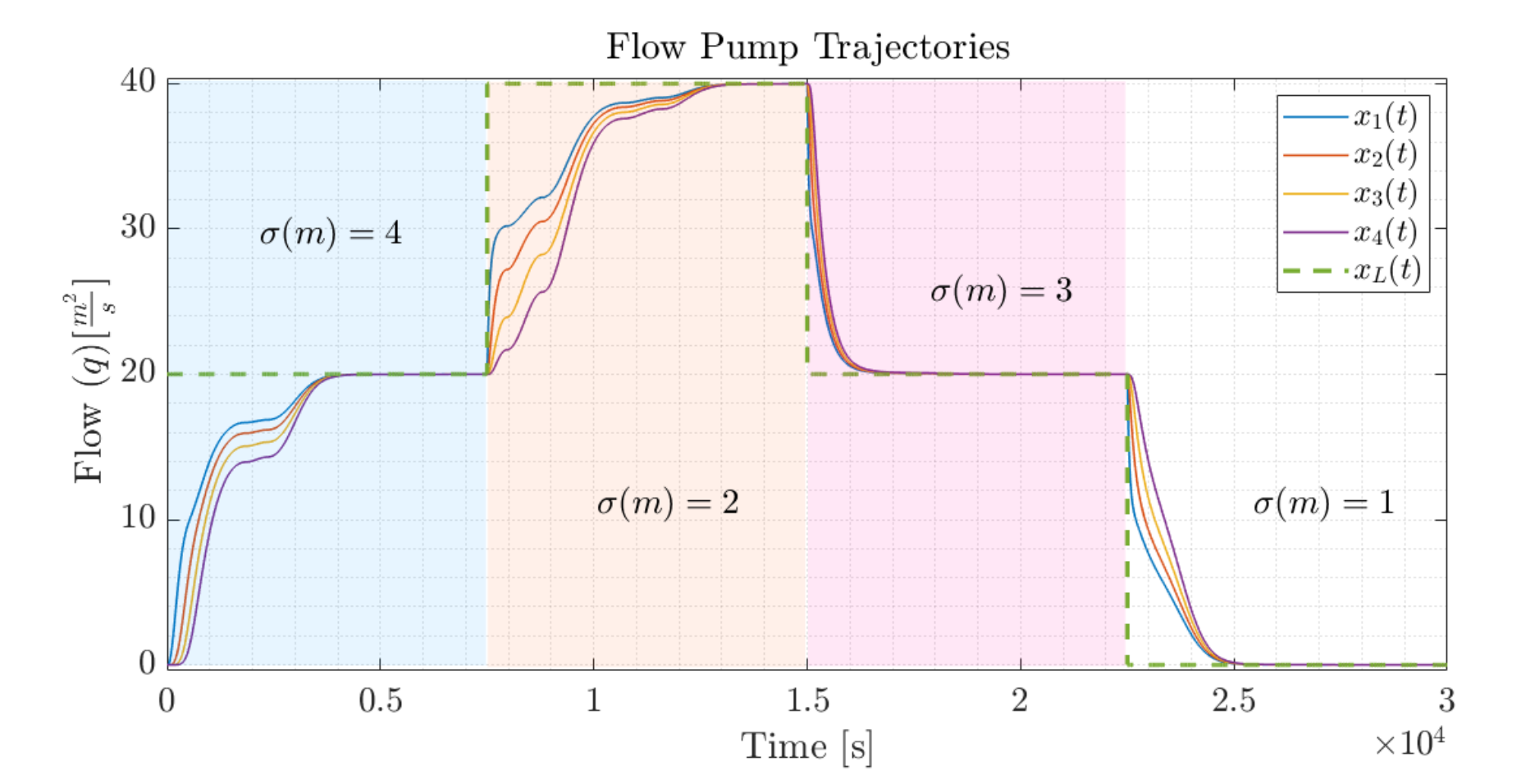}}
\caption{Pump flows using the switching algorithm in \eqref{eq:index_for_switching}.}
\label{Fig:Pumping System}
\end{figure}

In Figure~\ref{Fig:Pumping System}, we show the behavior of the set of pumps following a leader while the number of allowed followers changes with time. It is observed that regardless of the number of active agents ($\sigma$), consensus is reached. However, when $\sigma$ changes, the dynamics change. It is worth noting that, as intuitively expected, the convergence time is faster when $\sigma= 4$  than when  $\sigma= 2$. The intuition for that comes from the fact that when $\sigma= 4$, more followers are active per sample time and $p(m)=1$.  The same result, however, is not observed when  $\sigma= 3$, where the convergence time is even faster than when $\sigma = 4$. This behavior can be justified since after two iterations of $k$, all the followers would be active. Also, unlike the case, when $\sigma= 4$, the leader needs to change the follower it connects to, causing a faster convergence.

\section{CONCLUSIONS}

In this paper, we proposed a switching algorithm to control a hierarchical multi-agent system with a time-varying set of active agents. As expressed in Theorem~\ref{prop:convergence}, we observed that all followers could reach consensus regardless of the number of active agents. The implementation of the developed consensus algorithm on the application inspired examples, a MMC circuit and a water pumping system demonstrated the effectiveness and performance of the algorithm. As future work, it is still an open question whether we can improve the performance of the transient dynamics and achieve finite-time consensus to deal with a piece-wise constant leader state.


\bibliography{literature.bib}

\begin{thebibliography}{15}
\providecommand{\natexlab}[1]{#1}
\providecommand{\url}[1]{\texttt{#1}}
\providecommand{\urlprefix}{URL }
\expandafter\ifx\csname urlstyle\endcsname\relax
  \providecommand{\doi}[1]{doi:\discretionary{}{}{}#1}\else
  \providecommand{\doi}{doi:\discretionary{}{}{}\begingroup
  \urlstyle{rm}\Url}\fi

\bibitem[{Aguilar(2022)}]{Aguilar2022}
Aguilar, C.O. (2022).
\newblock \emph{An Introduction to Real Analysis}, volume~1.
\newblock Department of Mathematics, State University of New York.

\bibitem[{Cheng et~al.(2015)Cheng, Fan, and Zhang}]{CHENG2015112}
Cheng, Z., Fan, M.C., and Zhang, H.T. (2015).
\newblock Distributed mpc based consensus for single-integrator multi-agent
  systems.
\newblock \emph{ISA Transactions}, 58, 112--120.

\bibitem[{Jadbabaie et~al.(2003)Jadbabaie, Lin, and Morse}]{Jadbabaie2003}
Jadbabaie, A., Lin, J., and Morse, A. (2003).
\newblock Coordination of groups of mobile autonomous agents using nearest
  neighbor rules.
\newblock \emph{IEEE Transactions on Automatic Control}, 48(6), 988--1001.

\bibitem[{Kusumawardana et~al.(2019)Kusumawardana, Habibi, Wibawanto,
  Wicaksono, Prasetya, and Nurrahman}]{Kusumawardana2019}
Kusumawardana, A., Habibi, M.A., Wibawanto, S., Wicaksono, H., Prasetya, Y.,
  and Nurrahman, R. (2019).
\newblock Coordination power control of dc water pump system using dual-loop
  control and consensus algorithm.
\newblock In \emph{2019 ICEEIE}, volume~6, 37--42.

\bibitem[{Li and Zhang(2010)}]{Li2010}
Li, T. and Zhang, J.F. (2010).
\newblock Consensus conditions of multi-agent systems with time-varying
  topologies and stochastic communication noises.
\newblock \emph{IEEE Transactions on Automatic Control}, 55(9), 2043--2057.

\bibitem[{Lunze(2019)}]{Lunze2019}
Lunze, J. (2019).
\newblock \emph{Networked Control of Multi-Agent Systems}, volume~1.
\newblock Bookmundo, 1 edition.

\bibitem[{Mondal and Tsourdos(2022)}]{Mondal2022}
Mondal, S. and Tsourdos, A. (2022).
\newblock The consensus of non-linear agents under switching topology using
  dynamic inversion in the presence of communication noise and delay.
\newblock \emph{Proceedings of the Institution of Mechanical Engineers, Part G:
  Journal of Aerospace Engineering}, 236(2), 352--367.

\bibitem[{Morita and Ito(2018)}]{Morita2018}
Morita, R. and Ito, S. (2018).
\newblock Weak consensus with discrete-valued input and the performance
  dependency on its network topology.
\newblock In \emph{23rd International Symposium on Mathematical Theory of
  Networks and Systems}, 726--728.

\bibitem[{{Pereira Marca} et~al.(2021){Pereira Marca}, Duarte, Roes, and
  Wijnands}]{Ygor2021}
{Pereira Marca}, Y., Duarte, J., Roes, M., and Wijnands, C. (2021).
\newblock Extended operating region of modular multilevel converters using
  full-bridge sub-modules.
\newblock In \emph{23rd EPE ECCE Europe}. IEEE.

\bibitem[{Ren and Beard(2005)}]{Ren2005}
Ren, W. and Beard, R.W. (2005).
\newblock Consensus seeking in multiagent systems under dynamically changing
  interaction topologies.
\newblock \emph{IEEE Transactions on Automatic Control}, 50, 655--661.

\bibitem[{Shao et~al.(2018)Shao, Zheng, Huang, and Bishop}]{Shao2018}
Shao, J., Zheng, W.X., Huang, T.Z., and Bishop, A.N. (2018).
\newblock On leader–follower consensus with switching topologies: An analysis
  inspired by pigeon hierarchies.
\newblock \emph{IEEE Transactions on Automatic Control}, 63(10), 3588--3593.

\bibitem[{Shokri and Kebriaei(2020)}]{Shokri2020}
Shokri, M. and Kebriaei, H. (2020).
\newblock Leader–follower network aggregative game with stochastic agents’
  communication and activeness.
\newblock \emph{IEEE Transactions on Automatic Control}, 65(12), 5496--5502.

\bibitem[{Wei et~al.(2022)Wei, Kun, and Themistoklis}]{JIANG2022}
Wei, J., Kun, L., and Themistoklis, C. (2022).
\newblock Multi-agent consensus with heterogeneous time-varying input and
  communication delays in digraphs.
\newblock \emph{Automatica}, 135, 109950.

\bibitem[{Wolfowitz(1963)}]{wolfowitz1963}
Wolfowitz, J. (1963).
\newblock Products of indecomposable, aperiodic, stochastic matrices.
\newblock \emph{Proceedings of the American Mathematical Society}, 14(5),
  733--737.

\bibitem[{Yang et~al.(2011)Yang, Stoorvogel, and Saberi}]{Stoorvogel2011}
Yang, T., Stoorvogel, A., and Saberi, A. (2011).
\newblock Consensus for multi-agent systems — synchronization and regulation
  for complex networks.
\newblock In \emph{Proceedings of the American Control Conference, ACC 2011},
  5312--5317. IEEE Control Systems Society.

\end{thebibliography}
                                                   







\end{document}